# The role of the catalytic particle temperature gradient for SWNT growth from small particles

Feng Ding[*], Arne Rosén and Kim Bolton

Experimental Physics, School of Physics and Engineering Physics, Göteborg University and Chalmers University of Technology, SE-412 96, Göteborg, Sweden

**Abstract:** The Vapour-Liquid-Solid (VLS) model, which often includes a temperature gradient (TG) across the catalytic metal particle, is often used to describe the nucleation and growth of carbon nanostructures. Although the TG may be important for the growth of carbon species from large metal particles, molecular dynamics simulations show that it is not required for single-walled carbon nanotube growth from small catalytic particles.

## 1. Introduction

Although substantial progress has been made in the production of carbon nanotubes (CNTs) since their discovery more than a decade ago, the growth mechanism is still poorly understood. It is widely believed that the Vapour-Liquid-Solid (VLS) model, first used to explain the growth of Si nanowires [1] and then to describe the catalysed growth of carbon fibers (CFs) [2], provides a valid description of CNT growth [3]. According to the VLS model, carbon (from feedstock such as carbon-rich gases or graphite) dissolves into the liquid catalyst particle and, when the catalyst is supersaturated, carbon precipitates on the particle surface and nucleates the growth of CNTs.

*Corresponding author: Email-fengding@fy.chalmers.se
Tel- +46-31-7723294
On leave from Department of Physics, Qufu Normal University, Qufu, 273165, Shandong, P. R. China

When studying CF growth, Baker *et al.* proposed that a temperature gradient (TG) across the catalyst particle should be included in the VLS model [2]. The TG, illustrated in Fig. 1, arises since decomposition of feedstock occurs only on the part of the catalyst particle that is not covered by the growing CNT. If it is assumed that feedstock decomposition and dissolution of C atoms into the catalyst particle is exothermic and/or the precipitation of C atoms and their inclusion into the growing CNT is endothermic, then a TG arises between the hot region, where C atoms dissolve into the cluster, and the cold region, where C atoms precipitate [4]. This may be important for CNT growth for two reasons. First, the TG gives rise to a heat flow across the particle that can assist the transport of dissolved C atoms from the hot to the cold region. Second, saturation of, and precipitation from, the cold region requires a lower C concentration than the hot part. These two aspects lead to preferential C precipitation at the cold part of the particle where CNT growth occurs, hence preventing formation of graphene sheets (or other solid C structures) in the hot regions, which must remain uncovered to allow for continued decomposition of feedstock species (*i.e.*, it prevents poisoning of the catalyst particle).

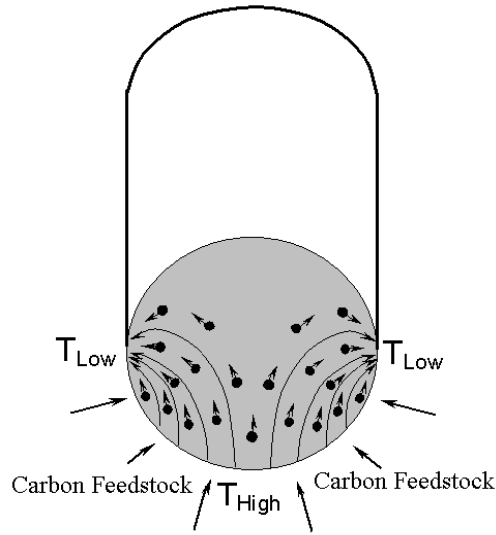

**Figure 1**. Illustration of the temperature gradient that can arise across a catalytic metal particle during CNT growth. The carbon flow (arrowed black dots) in the catalyst particle is driven by the temperature gradient (arrowed lines).

Although inclusion of a TG in the VLS model may be necessary to describe catalytic growth from large catalytic particles, it may not be important for the growth of narrow single-walled nanotubes (SWNTs) from small metal clusters [5]. For example, SWNTs are often grown from particles that have diameters of approximately 1 nm, and recent studies by D. Ciuparu *et al.* show that SWNTs can grow from cobalt clusters containing just 20-30 atoms [6]. It is unlikely that there is a temperature difference across such small particles since they have a high thermal conductivity and a small temperature difference would thus lead to an unphysically large heat flow [5]. Also, dehydrogenation of some hydrocarbons that are known to produce SWNTs, *e.g.*, methane, is endothermic [4] and, as discussed previously [7], this would not yield a TG in the correct direction. In addition, a recent thermodynamic calculation indicates that CNT growth is primarily driven by a concentration gradient (CG) and not a TG [8]. Thus, even though both a CG and a TG may be present under most experimental growth conditions, the CG may dominate and be sufficient for SWNT growth on small particles.

In this contribution we present results of molecular dynamics (MD) simulations of catalytic SWNT growth where it is seen that the VLS model provides a good description of the growth mechanism. However, SWNT growth does not require a TG, and growth persists even when a small TG is imposed across the metal particle but in the direction that opposes the growth process, *i.e.*, such that the SWNT grows from the hot region.

## 2. Potential energy surface and methods

Previous computational studies of catalysed SWNT growth have included static calculations [9], MD simulations based on analytic force fields [10] and density functional theory (DFT) based Car-Parrinello simulations [11]. While the Car-Parrinello approach has the advantage of obtaining forces directly from electronic structure theory, it suffers from the disadvantage that it is computationally very expensive. One can simulate many long trajectories (required for statistical analysis) when using MD based on analytic force fields, but it is important that the force field yields valid dynamics.

The MD simulations discussed here are based on an analytic potential energy surface (PES) that has been used to study the iron (Fe) catalysed nucleation of SWNTs [12]. Details of the PES and the SWNT nucleation mechanism appear elsewhere [12]. For the present contribution it is necessary to note that this PES yields the correct trends in the iron-carbide phase diagram (which is important if carbon saturation of the metal particle is important for CNT nucleation – as is the case for the VLS model) and the correct cluster melting point dependence on cluster size (which is important for determining the liquid/solid phase of the small particles used in SWNT growth) [13]. In addition, the PES distinguishes between the dissolved and precipitated C atoms, which is justified since dissolved C atoms are surrounded by Fe atoms and interact weakly, whereas precipitated C atoms interact strongly and form the SWNT. Also, two types of precipitated C atoms are modelled in the PES. These are the bond-saturated carbon atoms ($C_S$) that are bonded to three other C atoms (and that form part of a graphitic island or SWNT) and bond-unsaturated carbon atoms ($C_{Uns}$) that are bonded to at most two other C atoms (and are found, for example, at the edge/end of a graphitic island or SWNT). The $C_S$-Fe and $C_{Uns}$-Fe interaction energies were fit to results of DFT calculations, which show that the $C_S$-Fe bond energy (0.14 eV [14]) is about an order of magnitude weaker than the $C_{Uns}$-Fe bond energy (1.5 eV [15]).

Two types of trajectory conditions were considered in this study. The first mimics experimental chemical vapour deposition (CVD) conditions. These trajectories were initialised by thermalising pure $Fe_{50}$ cluster (50 iron atoms or a cluster diameter of ~1 nm) to 1000 K, and C atoms were subsequently added at a rate of one every 40 ps into the $Fe_{50}$ cluster. Adding C atoms to the cluster avoids the need to simulate catalytic decomposition of the carbon feedstock, which would require including the time-development of the electron density in the PES. Hence this work focuses on the role of the metal particle as a solvent, which is the important role in the VLS description. It should be noted that, since the small $Fe_{50}$ (and iron-carbide) clusters are in the liquid state at 1000 K, C diffusion in the cluster is rapid and the results presented here are thus insensitive to the region of the cluster where the C atoms are added. Also, the addition rate, which is more than 10000 times faster than experiment, is required to obtain SWNT nucleation in a tractable computational time. Decreasing the rate of addition does not affect the growth mechanism discussed here.

The second type of trajectory conditions are artificial, *i.e.*, they do not mimic realistic experimental conditions, and are chosen to test the effect of an imposed TG across the metal particle. The initial structures for these trajectories were partially grown SWNTs (*e.g.*, see Fig. 2c) and were obtained from the first type of trajectory conditions. During the propagation of these trajectories the temperature of the 25 Fe atoms nearest the growing SWNT was maintained at 1100 K, and that of the other atoms was kept at 1090 K. Due to the small size of the particle, this 10 K difference yields a significant TG, and it is in the opposite direction to that often included in the VLS model, that is, it opposes continued SWNT growth. In addition, C atoms were added to the cold region of the particle, away from the growing SWNT. These simulations are thus testing if SWNT growth will persist in spite of a TG that opposes diffusion of C to the SWNT end.

## 3. Results and discussion

Fig. 2 shows some snap shots during the nucleation of a typical SWNT structure (under simulation conditions that mimic CVD growth), which occurs via several steps: i) Carbon dissolves into the catalyst particle, ii) isolated carbon atoms precipitate on the cluster surface, iii) small carbon strings and polygons nucleate on the particle surface (Fig. 2a), iv) the strings and polygons grow into small carbon islands, v) typically one of the islands lifts off the surface to form a carbon cap while the other, smaller islands remain on the surface (Fig. 2b), vi) the smaller islands dissolve back into the cluster (Fig. 2b→d) and the cap increases in diameter and length to form the

SWNT.

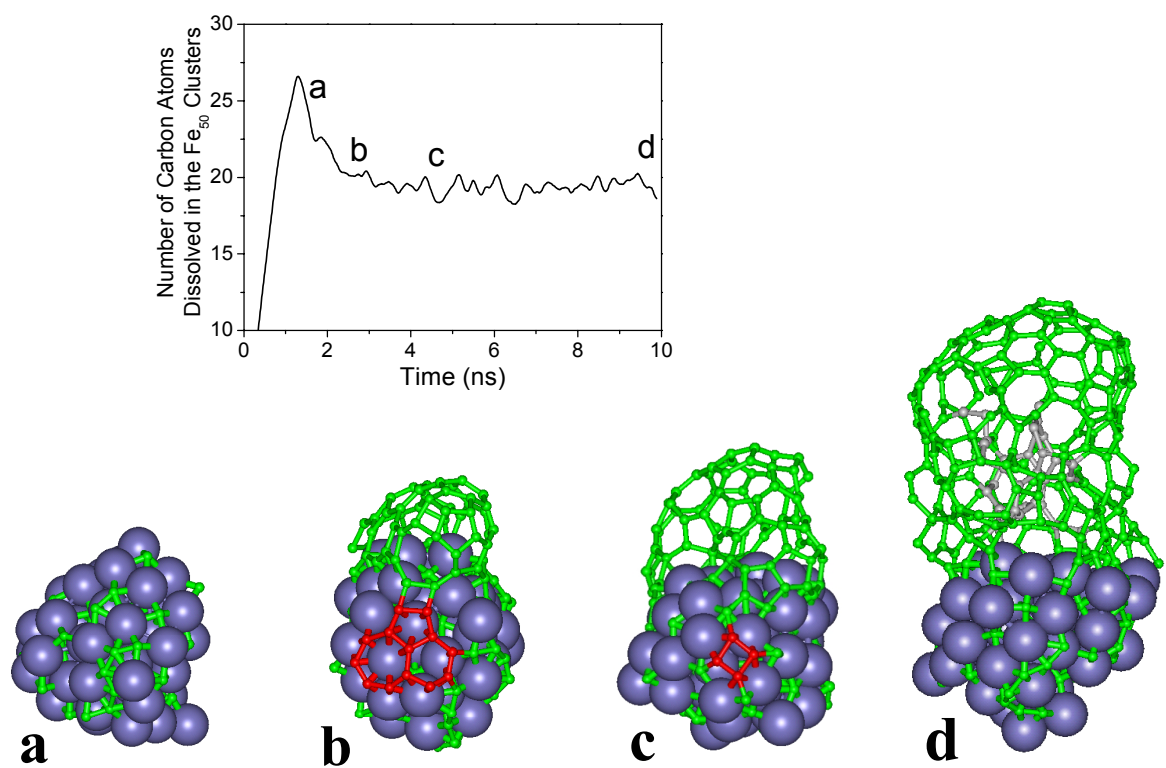

**Figure 2.** Nucleation and growth of a SWNT on the surface of an iron carbide cluster at 1000 K. The Fe atoms are shown as large spheres and the C atoms as a smaller ball-and-stick representation. The red C atoms form a small graphitic island that dissolves into the cluster and the grey atoms are joined to the inside of the nanotube due to defects. The insert shows the dissolved carbon content in the cluster during the growth (points a-d in the insert refer to the structure in panels a-d).

As discussed above, the primary reason for introducing a TG in the VLS model is to explain why dissolved C atoms diffuse to the growing SWNT instead of precipitating over the entire cluster surface, which would lead to encapsulation and poisoning of the particle. However, in the simulated growth mechanism shown in Fig. 2, two islands form on the surface and, in spite of this, only one island grows into a SWNT. In addition, once the SWNT begins to grow (Fig. 2c and d) no more islands form on the surface. This process, which is critical for the continued growth of the SWNT (and was previously explained in terms of the TG) is understood with reference to the insert in Fig. 2, which shows the dissolved carbon content in the iron carbide cluster as a function of time. These data are from the same trajectory shown in the figure, and structures a-d are shown on the graph in the insert. It is evident that nucleation sites form on the surface only when the iron carbide cluster is highly supersaturated in carbon (point a). Once these sites have formed, they nucleate larger islands, and the dissolved C concentration decreases to a 'saturated' level (the decrease between points a and b). Once one of the islands forms a cap (point b) and continues to grow (c and d) the carbon content remains at a constant, saturated level. This means that new islands cannot nucleate (since supersaturated conditions are required for nucleation). In this way only a single nanotube can grow from the particle and formation of new islands, that would poison the particle and stop SWNT growth, is not possible. Hence, for small particles such as the one studied here, changes in the dissolved carbon content is sufficient to explain continued growth of SWNTs, and TGs are not required.

In order to explicitly study the importance of the TG, a second simulation, which starts with the SWNT cap in Fig. 3a, was propagated such that a TG was artificially imposed over the catalyst cluster. In particular, a TG that opposes the SWNT growth was imposed by keeping the temperature in the region that is covered by the growing SWNT (1100 K) higher than the temperature in the region of the particle that is exposed (1090 K). In addition, as illustrated in Fig. 3, C atoms are added into the cold region of the particle, and thus have to diffuse from cold to hot regions for the SWNT growth to persist. In spite of the opposing TG, SWNT growth persists so that a longer SWNT (Fig. 3b) is obtained after 6 ns, during which time 75 C atoms have added to the SWNT.

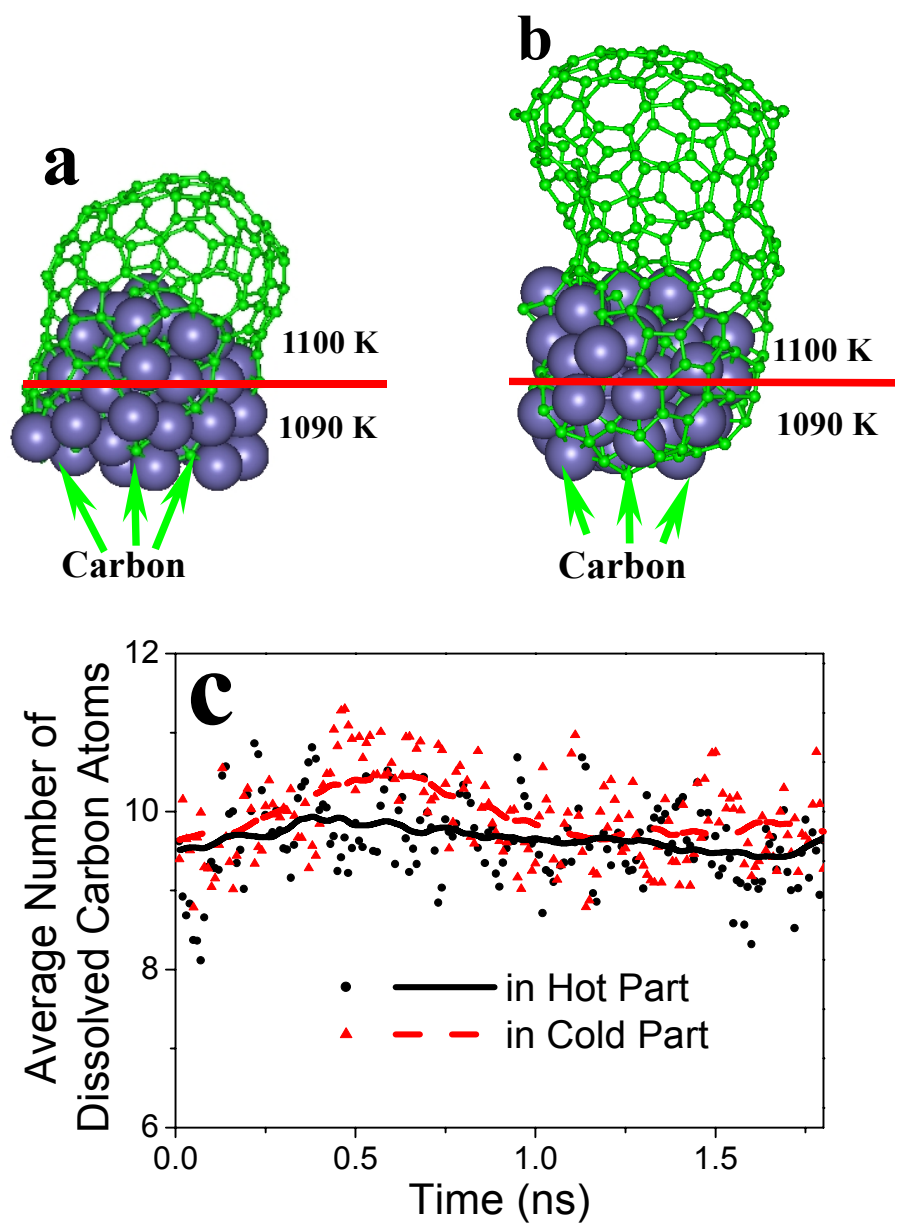

Figure 3. SWNT growth (a to b) persists in spite of an artificially induced TG that opposes the nanotube growth. During this simulation carbon atoms are added to the lower, colder part of the cluster. The average number of dissolved carbon atoms in the hot and cold parts of the cluster are shown in c. The scatter symbols are obtained from averaging over 10 ps intervals, and the lines from averaging over 200 ps.

The above results reveal that a TG is not required for continued SWNT growth from small metal particles. An alternative explanation, as suggested by others (*e.g.*, [8, 16, 17]), is that a concentration gradient (CG) is sufficient to describe the nucleation and growth of SWNTs on small catalyst particles. Fig 3c shows that the dissolved carbon concentration (points are obtained from 10 ps averages and the lines from 200 ps averages) in the cold part of the catalyst cluster is slightly higher than that in the hot part, and that one can have an inversion of the concentration gradient (CG) at certain times along the trajectory (*i.e.*, as seen in the figure the C content in the hot region can be larger than that in the cold region). Hence, the simulations are consistent with models of CG-driven SWNT growth, but even though the small concentration difference shown in the figure results in a large CG (due to the 1 nm size of the catalyst particle), the simulations performed here do not provide a detailed understanding of the CG-driven growth mechanism.

CG-driven SWNT growth is illustrated in Fig. 4, where regions of the cluster that are near to the open SWNT end have low C concentrations (since C is removed when it precipitates and forms part of the SWNT), and regions of high C concentration far from the SWNT. C in the high concentration regions is continuously replenished by catalytic decomposition of feedstock (on the surface of the cluster that is not covered by the SWNT) and, due to the CG, flows to the end of the SWNT. The MD results presented here show that, in spite of the higher C concentration in the exposed region of the particle, the C does not precipitate in this region since the cluster is never highly supersaturated during SWNT growth. (In fact, simulations show that an extremely high rate of C addition to the particle can lead to highly supersaturated conditions in the exposed region which results in precipitation and poisoning of the particle.) It may be noted that Fig. 4 shows (equally) large C densities in all regions of the particle that are not connected to the open SWNT end, which is the case if C diffusion in the particle is far faster than SWNT growth. If diffusion is slower than SWNT growth (*e.g.*, high feedstock pressures) then the part of the cluster that is accessible to feedstock gases (*i.e.*, the lower part of the cluster in Fig. 4) will have a higher C concentration than the region that is covered by the SWNT.

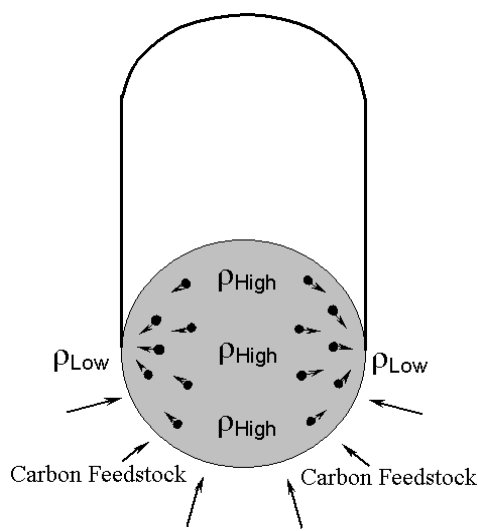

**Figure 4**. Illustration of the concentration gradient (CG) across the catalytic metal particle during SWNT growth. Incorporation of atoms into the SWNT structure lowers the C concentration near the open SWNT end.

## 4. Conclusion

The MD simulations show that catalytic SWNT nucleation and growth is well described by the VLS model. A temperature gradient is not required for SWNT growth on the small particles studied here. Instead, the edges of the graphitic cap – or the ends of the SWNT – act as a C sink during the growth process. This reduces the carbon concentration in the regions of the particle that are near the cap edge (or SWNT end), which leads to the diffusion of C atoms from regions of high C concentration to the cap or SWNT edge, where they incorporate into the growing structure.

**Acknowledgements**

We are grateful to the Swedish Foundation for Strategic Research (CARAMEL consortium) and the Swedish Research Council for financial support, and for time allocated on the Swedish National Supercomputing facilities.